\documentclass[]{raa}    
\usepackage{graphicx,times}             
\usepackage{natbib}
\usepackage{amssymb,amsmath}
\usepackage{grffile}

\bibpunct{(}{)}{;}{a}{}{,}

\usepackage[a4paper=true,pagebackref=true]{hyperref}
\hypersetup{colorlinks = true, linkcolor = blue, anchorcolor = red, citecolor = blue, filecolor = red, pagecolor = red, urlcolor = green}

\begin{document}

\title{Theoretical Analysis of Random Scattering Induced by Microlensing}

 \volnopage{Vol.0 (20xx) No.0, 000--000}      
 \setcounter{page}{1}          

\author{Wenwen Zheng \inst{1,2}, Hou-zun Chen \inst{1,2}, 
Xuechun Chen \inst{3,4},
Guoliang Li \inst{1},
}
  \institute{Purple Mountain Observatory, Chinese Academy of Sciences, Nanjing, Jiangsu, 210023, China; \\
    \and
   School of Astronomy and Space Science, University of Science and Technology of China, Hefei, Anhui, 230026, China;\\
    \and  
   Institute for Frontier in Astronomy and Astrophysics, Beijing Normal University, Beijing, 102206, China;\\
    \and
    Department of Astronomy, Beijing Normal University, Beijing, 100875, China;\\
    {\it Emails: wwzheng@pmo.ac.cn}\\
\vs\no
   {\small Received~~20xx month day; accepted~~20xx~~month day}
}



\abstract{Theoretical investigations into the  deflection angle caused by microlenses offer a direct path to uncovering principles of the cosmological microlensing effect.
This work specifically concentrates on the the probability density function (PDF) of the light deflection angle induced by microlenses. We have made several significant improvements to the widely used formula from \cite{1986ApJ...306....2K}.Firstly, we update the coefficient from 3.05 to 1.454, resulting in a better fit between the theoretical PDF and  our simulation results.  
Secondly, we developed an elegant fitting formula for the PDF that can replace its integral representation within a certain accuracy, which is numerically divergent unless arbitrary upper limits are chosen.
Thirdly, To facilitate further theoretical work in this area, we have identified a more suitable Gaussian approximation for the fitting formula. 
}



   \authorrunning{Wenwen Zheng et al.}            
   \titlerunning{Theoretical Analysis of Random Scattering Induced by Microlensing}  

   \maketitle

\section{Introduction}
The strong lensing effect caused by a foreground galaxy may produce multiple macroimages, and in the mean time, the compact objects inhabiting the galaxy can further deflect the light rays, causing a macroimage to split into microimages. This phenomenon is known as the cosmological microlensing effect and was described by \cite{1979Natur.282..561C}. 
Although it is extremely difficult to directly distinguish microimages, the flux anomaly between macroimages can provide a clue, because light rays from different macroimages are affected by different distributions of microlenses.
It is possible to detect this effect in a time-domain observation: The relative motion between the background source, the lens galaxy, the randomly moving stars and the observer induces differences between light curves of macroimages (when other impacts are deducted), which was first confirmed in quasar system Q2237+0305 presented by \cite{1989AJ.....98.1989I}.
Based on the theory of microlensing, the variation of light curves can contain information on both the source size and the lens mass distributions. 
This has led to a series of pioneering research efforts on constraining the inner structure of quasars \citep{2004ApJ...605...58K,2005ApJ...628..594M,2012A&A...544A..62S} and the mass distribution of compact objects in the lens galaxy \citep{2001MNRAS.320...21W,2009ApJ...706.1451M,2014ApJ...793...96S,2019ApJ...885...75J}.These research efforts have quickly produced significant results.
In recent years, a host of unprecedented events has been observed, where individual stars at cosmological distances are  magnified to an extreme degree due to their close proximity to massive cluster caustics \citep{2018NatAs...2..334K, 2018NatAs...2..324R, 2019ApJ...881....8C, 2019ApJ...880...58K, 2022A&A...665A.134D, 2022Natur.603..815W}.
 The smooth macrocaustic will be broken into a net of microcaustics in the presence of intra-cluster light (ICL) in clusters, as described in \cite{2017ApJ...850...49V}, inducing a variety of fluctuations in the magnification of a star as it moves across the micro-caustics. 
Similar to lensed quasar system, microlensing light curves also provide the possibility of digging information about the size of the source star, the fractions of ICL, and even MACHOS.
Therefore, extracting the information contained in the light curves is essential in microlensing studies, and simulations have become the most commonly used method thanks to the vigorous development of computer technology.

Several approaches have been developed for generating microlensing light curves, with the most commonly used method being the inverse ray shooting (IRS) \citep{1986A&A...166...36K,1986MPARp.234.....S}. Since then, various optimized versions of the IRS method have been proposed \citep{1999JCoAM.109..353W,  2006ApJ...653..942M, 2010NewA...15...16T, 2021A&A...653A.121S, 2022ApJ...931..114Z}. 
They share a basic procedure: light rays are traced back from the image plane to the source plane to generate a magnification map, which is then convolved with a moving source to produce a set of light curves. 
However, the processing of huge amounts of data still poses a great challenge for standard simulations when dealing with caustic-crossing events.
New methods, such as those proposed by \cite{2022MNRAS.514.2545M} and \cite{2022A&A...665A.127D}, have been developed to reduce computational costs while still achieving accurate light curves. Additionally, for point sources, there are other methods \citep{1993MNRAS.261..647L,1993ApJ...403..530W} that are designed to directly produce individual light curves without generating a whole magnification map.

Nonetheless, all of these approaches face a common challenge: the presence of compact objects can cause some of the light rays to be deflected out of the source plane, leading to a loss of simulation accuracy. To address this issue, the size of the shooting area needs to be increased when setting the image plane in order to ensure precision\citep{ 1986MPARp.234.....S, 1999JCoAM.109..353W}. 
Rewritten: It is important to note that the size of the increased shooting area should be determined based on the distribution of the microlens-induced deflection angle.
This distribution is directly related to the magnification and observable variation in the lensed source light curves, which can provide valuable insight for future theoretical research \citep{2021arXiv210412009D}.
Therefore, further study on microlensing deflection angle is necessary.

Remarkable results were presented in \cite{1986ApJ...306....2K} (hereafter K86), where the integral form of the probability density distribution of the deflection angles is given in detail (their Equation (13)). 
 K86 concluded that the probability density function (PDF) has a Gaussian form at small deflection angles and is inversely proportional to the quartic of deflection angle at the infinite end. This provides a basis for all types of methods that include an ``image-to-source-plane ray mapping" part to set a buffer to control missing light rays.

In our previous simulation work \citep{2022ApJ...931..114Z}, we included a ``negative mass sheet'' term in the deflection angle equation, Which led us to reanalyze the PDF of the deflection angles. Our new analysis yielded a different result that showed better consistency with our simulation. 
This paper is organized as follows:  In Section~\ref{sec:PDF of the deflection angle}, we present the theoretical derivation of the PDF of the deflection angle.
This section is divided into three subsections:  In subsection~\ref{subsec: the negative mass sheet} we re-deduced the PDF of the deflection angle.  Subsection~\ref{subsec:the fitting formula} describes the process of obtaining a fitting formula for the PDF and in subsection~\ref{sec:Gaussian approximation} we present the methods we used to obtain a Gaussian approximation.
A short summary of this work is provided in section~\ref{sec:Summary}.

    
\section{Probability density function of the deflection angle} \label{sec:PDF of the deflection angle}
Ideally, in the absence of compact objects, all light rays traveling from an image rectangle with a side length ratio of $\mid1-\kappa-\gamma\mid$ / $\mid1-\kappa+\gamma\mid$ would fall within a square.
 However, the presence of compact objects causes some of the rays to scatter out, indicating the effect of the deflection angle induced by microlensing.
Referring to our previous work \cite{2022ApJ...931..114Z}, the general form of the deflection angle can be written as
\begin{equation}
{\pmb \alpha(\pmb \theta)} =  \left( 
\begin{array}{cc}
\kappa+\gamma & 0 \\
0 & \kappa-\gamma 
\end{array}
\right) {\pmb \theta}
+\sum_{i=1}^{N_*} m_i \frac{({\pmb \theta} - {\pmb \theta}_i ) }
{\vert{\pmb \theta} - {\pmb \theta}_i\vert^2}
+\pmb \alpha_{-\kappa_*},
\label{eqn:alpha1}
\end{equation}
where $\pmb\theta$ and $\pmb\theta_i$ denote the light positions and lens positions on the image/lens plane, respectively. $N_*$ is the number of microlenses (hereafter stars, as they are the majority in number) while $m_i$ denotes their mass. $\kappa$ is the dimensionless surface mass density of the mass sheet (which contains both the compact and smooth mass) and $\gamma$ is the external shear; both derived from the strong lens model.

The general form of the deflection angle can be written as a sum of three terms. The first term represents the impact of local convergence and shear. The second term indicates the contribution from the individual stars, with the mean surface mass density of $\kappa_*$, whereas the third term $\pmb\alpha_{-\kappa_*}$ represents a negative mass sheet that is only composed of the smooth matter, with the same $\kappa_*$ value to cancel out the redundant mass from the second term.
Thus, the sum of the second and the third terms represents the perturbation caused by compact objects. This perturbation will be the deflection angle we mainly focus on in this work.

\subsection{PDF of the deflection angle} \label{subsec: the negative mass sheet}

In K86, the integral representation of scattering PDF has the form (their Equation (13)),
\begin{equation}
\small
 \rho_N(\phi) = \frac{1}{2\pi\phi_0^2} \int_{0}^{\infty} x\mathrm{d}x\mathrm{J}_0\left ( \frac{x\phi}{\phi_0} \right ) \exp\left [ - \frac{x^2}{2}\ln_{}{\left (\frac{3.05N^{1/2}}{x}\right )}\right ],
 \label{eqn:Katz}
\end{equation}
where $\phi$ denotes the deflection angle, and $\mathrm{J}_0$ is the Bessel function of the first kind.
$x = \phi/\phi_0$ and $\phi_0$ is a scaling parameter that is related to the number of microlenses $N$, as will be explained in detail later in the text.

However, in their approach, they only considered the deflection angle contributed from the individual stars, so  the negative mass sheet was taken as an overall variable. Consequently, in every situation, the negative mass sheet  occupied by each star will not overlap.
In contrast, we assume that each star carries a negative circular mass sheet with radius $R$,  and that the two parts of each star have equal and opposite masses, resulting in a net zero mass. 
This means that both parts act as a scattering source when calculating the deflection angle. 
In this situation, the negative mass sheets of the scattering sources may overlap depending on their distance, which we consider to be a more natural scenario for random scattering of microlensing.

We start by defining $\phi_{\mathrm{min}}$ in the same way as K86 did. This quantity represents the minimum deflection angle $\phi_{\mathrm{min}} = 4GM/c^2R$, induced by a point mass $M$ when light rays are distributed in a circle centered on it, with a radius of $R$.

Then for one scattering source, the deflection angle at impact parameter $b$ ($b\le$ $R$) is 
\begin{equation}
    \phi(b)=\frac{4GM}{c^2b} -\frac{4G (M\cdot\frac{b^2}{R^2})}{c^2b}=\phi_{\mathrm{min}}\frac{R}{b}\left ( 1-\frac{b^2}{R^2}\right ).
\label{eqn:phi}
\end{equation}
As disturbances with deflection angle greater than $\phi(b)$ occur within radius $b$,  the cumulative probability would be $P(> \phi)=\frac{b^2}{R^2}$, and combining it with Equation~(\ref{eqn:phi}) we have
\begin{equation}
    P^2-(2+x^2)P+1=0, \;\;\;\mathrm{where} \; x=\phi/\phi_{\mathrm{min}}.
\end{equation}
 Upper and lower bounds of the equation can be derived when $x$ tends to zero and infinity, conforming to the property of cumulative probability
\begin{equation}
    0\le P=1+\frac{x^2-\sqrt{x^4+4x^2}}{2}\le 1.
\end{equation}

\begin{figure*}
	
	\centering
	\includegraphics[width=10cm]{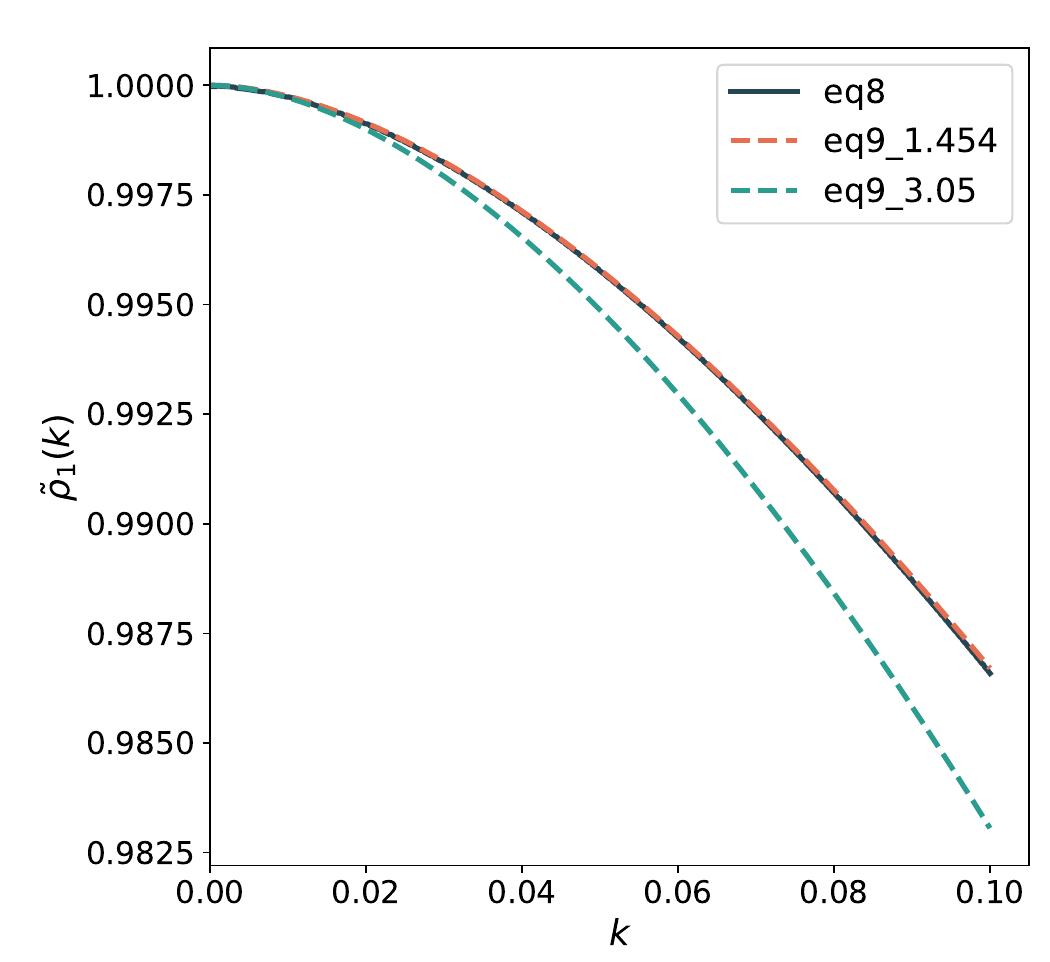}
    \caption{The comparison of Equation~(\ref{eqn:tilde_rho1_k}) and Equation~(\ref{eqn:tilde_rho1_k_2}). The dark blue solid line represents the numerical solution result of Equation~(\ref{eqn:tilde_rho1_k}), while orange and green dashed lines correspond to Equation~(\ref{eqn:tilde_rho1_k_2}) and the result of K86, respectively.
    }
    \label{fig:compare_eq8_eq9}
\end{figure*}

Accordingly, the probability density turns out to be
\begin{equation}
    p_1(\phi) = -{P}'(>x)\frac{\mathrm{d} x}{\mathrm{d} \phi}=\frac{1}{\phi_{\mathrm{min}}}\left [ \frac{x^3+2x}{\sqrt{x^4+4x^2}}-x\right ].     
\end{equation}
The two-dimensional probability is
\begin{equation}
    \rho_1(\phi)=\frac{p_1(\phi)}{2\pi\phi}=\frac{1}{2\pi\phi_{\mathrm{min}}^2}\left ( \frac{x^2+2}{\sqrt{x^4+4x^2}}-1\right).
\end{equation}

Taking the Fourier transformation of $\rho_1(\phi)$, we find
\begin{equation}
\begin{aligned}
    \tilde\rho_1(k)= &2\pi\int_{0}^{\infty}\rho_1(\phi)\mathrm{J}_0(k\phi)\phi\mathrm{d}\phi\\
                   = &\frac{1}{\phi^2_{\mathrm{min}}} \int_{0}^{\infty} \mathrm{J}_0(k\phi)\left ( \frac{x^2+2}{\sqrt{x^4+4x^2}}-1\right)\phi\mathrm{d}\phi\\
                   = &\int_{0}^{\infty} \mathrm{J}_0(ax)\left ( \frac{x^2+2}{\sqrt{x^2+4}}-x\right)\mathrm{d}x,
\end{aligned}
\label{eqn:tilde_rho1_k}
\end{equation}
where $a=k\phi_{\mathrm{min}}$.

Through several derivations, we derive
\begin{equation}
\begin{aligned} 
    \tilde\rho_1(k) \simeq& 1-\frac{a^2}{2} \ln \left( \frac{1.866}{a\sqrt[4]{e}} \right)
    \simeq  1-\frac{a^2}{2} \ln \left(\frac{1.454}{a}\right) \\
    \simeq&  \exp\left [ -\frac{a^2}{2}\ln\left(\frac{1.454}{a}\right)\right]\;\; (\mathrm{when} \; a \ll 1),
    \label{eqn:tilde_rho1_k_2}
\end{aligned}
\end{equation}
and for more details on this part of the derivation please refer to Appendix A.

Equation~(\ref{eqn:tilde_rho1_k_2}) has the same form as the one derived by K86,  which has been modified from 3.05 to 1.45. Fig. ~\ref{fig:compare_eq8_eq9} displays a comparison between Equation~(\ref{eqn:tilde_rho1_k}) and ~(\ref{eqn:tilde_rho1_k_2}), along with the deviations caused by the different numerators.
 Here we find that Equation~(\ref{eqn:tilde_rho1_k}) and ~(\ref{eqn:tilde_rho1_k_2}) are in good agreement, which indicates that the approximations used are reasonable. On the other hand, the difference brought by the two numerators is non-negligible.

\begin{figure*}
	\includegraphics[width=\linewidth]{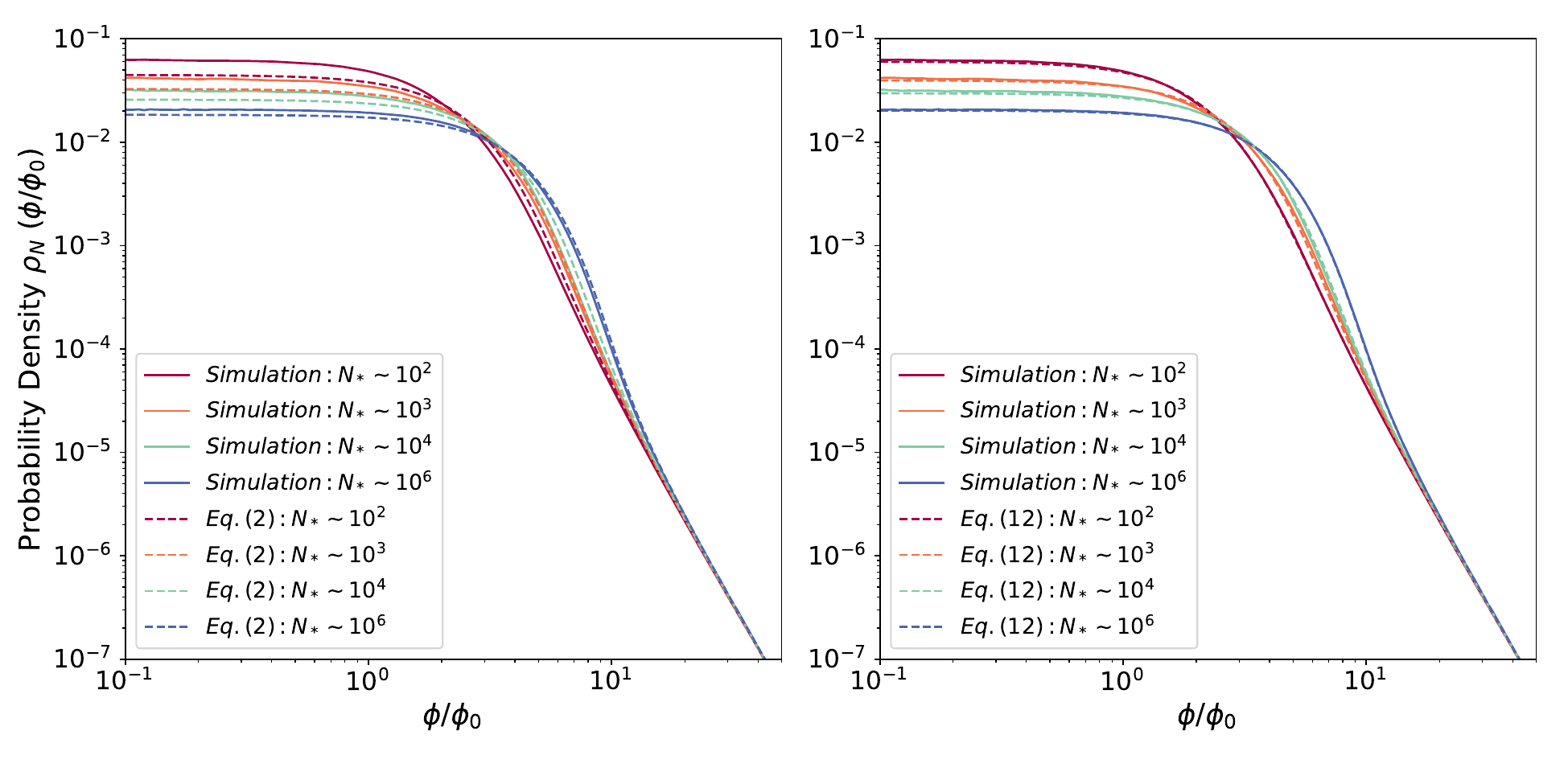}
    \caption{The probability density distribution of the deflection angle. Solid lines represent the distribution realized by our simulation, dashed lines represent the numerical results of Equation~(\ref{eqn:Katz}) concluded by K86 (left), and the newly acquired Equation~(\ref{eqn:neweq13}) (right). The star mass is uniformly distributed with $M = 1M_{\odot}$, and different colors stand for deflection angle distributions with different numbers of stars: $10^2$, $10^3$, $10^4$ and $10^6$. 
    }
    \label{fig:simulation_numerical_eq13}
\end{figure*}

Suppose there are $N$ stars with uniform mass randomly distributed in a circular plane with radius $R$. We choose a circular plane, rather than a rectangular one, for ease of calculation, which will not lead to a different result. Each star and its corresponding negative mass sheet occupy a circular area with radius $R_0=R/\sqrt{N}$.

Hence, the final scattering angle distribution is to repeat the convolution operation of $\rho_1(\phi)$ $N$ times. In Fourier space we have
\begin{equation}
\tilde\rho_N(k)=\tilde\rho_1^N(k)\simeq \exp\left [ -\frac{Na^2}{2}\ln\left({\frac{1.454}{a} }\right)\right],
\label{eqn:tilde_rhoN_k}
\end{equation}
then transform $\tilde\rho_N(k)$ back to real space 
\begin{equation}
\small
 \rho_N(\phi) = \frac{1}{2\pi} \int_{0}^{\infty} \mathrm{J}_0({k\phi}) \exp\left [ - \frac{Nk^2\phi_{\mathrm{min}}^2}{2}\ln_{}{\left (\frac{1.454}{k\phi_{\mathrm{min}}}\right )}\right ]k\mathrm{d}k.
 \label{eqn:neweq10}
\end{equation}

We follow K86 to define $\phi_0=\sqrt{N}\phi_{\mathrm{min}}$ as a scaling parameter, which denotes the deflection angle for a light ray passing by a point mass $M$ at a radius of $R_0$, to rewrite the above equation as
\begin{equation}
\small
 \rho_N(\phi) = \frac{1}{2\pi\phi_0^2} \int_{0}^{\infty} \mathrm{J}_0\left ( \frac{x\phi}{\phi_0} \right ) \exp\left [ - \frac{x^2}{2}\ln_{}{\left (\frac{1.454\sqrt{N}}{x}\right )}\right ]x\mathrm{d}x.
 \label{eqn:neweq13}
\end{equation}
This is the integral representation of the scattering PDF in terms of our negative mass sheet model. Not surprisingly, it has the same form as Equation~(\ref{eqn:Katz}) except for the coefficient 1.454.

In Fig.~\ref{fig:simulation_numerical_eq13} we compare the numerical results (dashed lines) of Equation~(\ref{eqn:neweq13}) and 
 ~(\ref{eqn:Katz}) (K86 Equation (13)) with a series of simulation results (solid lines) obtained using with our GPU-PMO code (\cite{2022ApJ...931..114Z}). 
The PDF of the deflection angle is calculated with a varying number of lenses (from $10^2$ to $10^7$) signified by different colors in Fig.~\ref{fig:simulation_numerical_eq13}.
Only $10^2$, $10^3$, $10^4$ and $10^6$ are shown in the figure for clarity. 
The numerical result for Equation ~(\ref{eqn:Katz}) is plotted in the left panel while that for Equation ~(\ref{eqn:neweq13}) in the right panel. 
The deflection angle is scaled by $\phi_0$.
As affirmed in Fig.~\ref{fig:simulation_numerical_eq13}, our modified equation Equation~(\ref{eqn:neweq13}), fits the simulation results better compared to Equation~(\ref{eqn:Katz}). This suggests that our assumption and derived new coefficient of 1.454 are reasonable.

\subsection{The fitting formula for deflection angle probability density function}  \label{subsec:the fitting formula}

As mentioned in K86, the limits of the integration of Equation~(\ref{eqn:Katz}) cannot run from 0 to $\infty$, otherwise it will lead to a formal divergence.
Therefore, in practice, K86 terminates the integration when the integrand becomes exponentially small. They calculate the asymptotic limits for Equation~(\ref{eqn:Katz}) and conclude that for small $\phi$, the density is a Gaussian, and for large $\phi$, it is inversely proportional to the fourth power of the deflection angle.

In this work, we attempt to obtain the analytic result of Equation~(\ref{eqn:neweq13}) rather than rely on the asymptotic limits.


First, we define $t=\phi/\phi_0$, replace the coefficient 1.454 with the parameter $\beta$, and define $\sigma^2=\mathrm{ln}(\beta\sqrt{N})$, to rewrite Equation~(\ref{eqn:neweq13})
\begin{equation}
\small
 \rho_N(t) = \frac{1}{2\pi} \int_{0}^{\infty} \mathrm{J}_0(xt) \exp\left [ - \frac{x^2}{2}(\sigma^2-\mathrm{ln}x ) \right]x\mathrm{d}x.
 \label{eqn:rhont with beta}
\end{equation}
By inserting Taylor expansion
\begin{equation}
\small
    \exp\left [ -\frac{x^2}{2}(\sigma^2-\ln x)\right]=\exp \left ( -\frac{\sigma^2x^2}{2}  \right )  \left ( 1+\frac{x^2}{2}\ln x+...\right ) 
\end{equation}
into Equation~(\ref{eqn:rhont with beta}), we derive
\begin{equation}
\begin{aligned}
    \rho_N(t) = &\frac{1}{2\pi} \int_{0}^{\infty} x\mathrm{J}_0(xt) \exp\left ( - \frac{\sigma^2x^2}{2} \right)\mathrm{d}x\\
    +&\frac{1}{4\pi}\int_{0}^{\infty}x^3\ln x\mathrm{J}_0(xt)\exp\left ( - \frac{\sigma^2x^2}{2} \right)\mathrm{d}x+... \\.
\label{eqn:rhont 2 integral}
\end{aligned}
\end{equation}

The specific process of solving these two integrals is described in Appendix B, here we jump to the result
\begin{equation}
\small
\begin{aligned}
\rho_N(t) = &\frac{1}{2\pi \sigma^2} \exp{ \left(-\frac{t^2}{2\sigma^2}\right)}\\
   &+\frac{1}{4\pi \sigma^4}\left[  \left ( \ln\frac{2}{\sigma^2}-\gamma  \right )\left ( 1-\frac{t^2}{2\sigma^2}\right )\exp \left( -\frac{t^2}{2\sigma^2}\right) \right.\\
   & \left. +\sum_{n=0}^{\infty}\frac{(n+1 )}{n!}H_{n+1}\left ( -\frac{t^2}{2\sigma^2}\right )^n \right],
\end{aligned}
\end{equation}
where the constant $\gamma=0.5772$... denotes the Euler-Mascheroni constant, and $H_n= {\textstyle \sum_{k}^{}}\frac{1}{k}$ is the harmonic number. For $z>0$ the latter infinite series has a closed form written as below
\begin{equation}
\begin{aligned}
    &\sum_{n=0}^{\infty}\frac{(n+1 )}{n!}H_{n+1}(-z)^n \\
    &= (1-z)e^{-z}(\gamma +\ln z-\mathrm{Ei}(z))+2e^{-z}-1,
\end{aligned}
\end{equation}
$\mathrm{Ei}(z)=\int_{-\infty }^{z}\frac{e^w}{w}\mathrm{d}w$ is the exponential integral function. 
Notably, the Taylor expansion of $\mathrm{Ei}(z)$ near the origin can be expressed as 
\begin{equation}
      \mathrm{Ei}(z)=\gamma +\ln z+\sum_{n=1}^{\infty }\frac{z^n}{n\cdot n!},
\label{eqn:Ei expansion origin}
\end{equation}
and for the infinite end, $\mathrm{Ei}(z)$ has an asymptotic expansion state as 
\begin{equation}
      \mathrm{Ei}(z)=\frac{e^z}{z}\left ( 1+\frac{1}{z}+\frac{2}{z^2}+\frac{6}{z^3}+\frac{24}{z^4}+...\right ),\;\;z\to \infty
\label{eqn:Ei expansion}
\end{equation}
Theoretically, series ~(\ref{eqn:Ei expansion}) does not converge for a fixed $z$, therefore, the asymptotic series needs to be cut off before the term becomes $z^n<n!$.

\begin{figure*}
	\includegraphics[width=\linewidth]{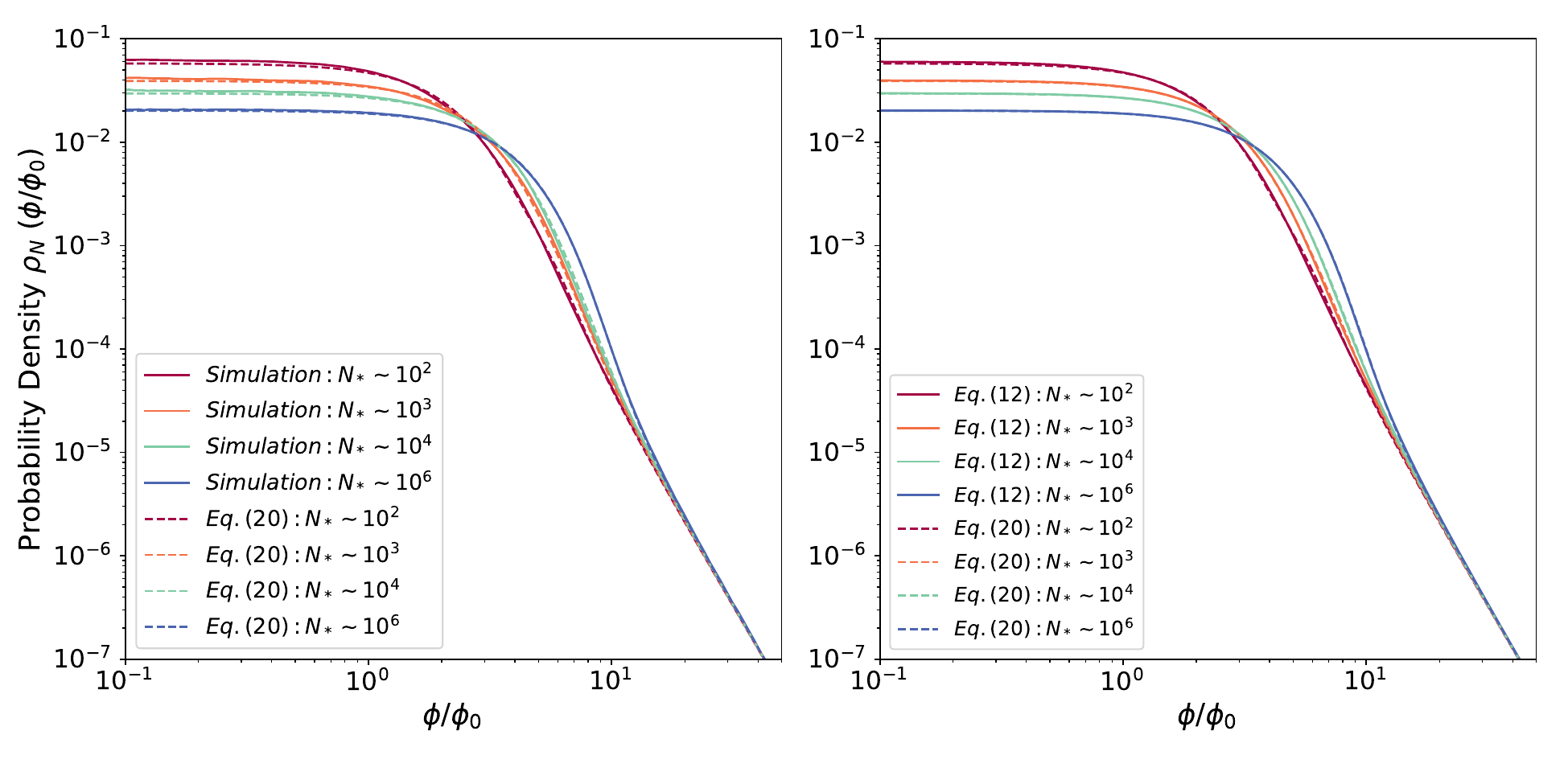}
    \caption{Comparison between the simulation (solid lines in left panel), Equation~(\ref{eqn:neweq13}) (solid lines in right panel) and Equation~(\ref{eqn:Final rhoN}) (dashed lines in both panels). Different colors stand for the deflection angle distributions with different numbers of stars: $10^2$, $10^3$, $10^4$ and 
 $10^6$, which are uniformly distributed with $\left \langle M \right \rangle = 1M_{\odot}$. 
    }
    \label{fig:simulation_neweq}
\end{figure*}

Altogether, we obtain the fitting formula of the PDF

\begin{equation}
\begin{aligned}
    \rho_N(t)\simeq &\frac{e^{-z}}{2\pi \sigma^2}+\frac{1}{4\pi \sigma^4}\left\{ 2e^{-z}-1 
     -(1-z)e^{-z}\left [ \mathrm{Ei}(z) -\ln z-\ln \frac{2}{\sigma^2} \right ] \right\},
     \;\;z=\frac{t^2}{2\sigma^2} \\
              =&   \frac{1}{2\pi \sigma^2}\exp{ \left(-\frac{t^2}{2\sigma^2}\right)} + \frac{1}{4\pi \sigma^4}\left \{ 2\exp{ \left(-\frac{t^2}{2\sigma^2}\right)}-1 \right. \\
              & -(1-\frac{t^2}{2\sigma^2})\exp{ \left(-\frac{t^2}{2\sigma^2}\right)} \left [ \mathrm{Ei}\left ( \frac{t^2}{2\sigma ^2} \right )-\ln \left ( \frac{t^2}{2\sigma^2}\right )  \left. -\ln \left ( \frac{2}{\sigma^2}\right ) \right ]  \right \}.
\end{aligned}
\label{eqn:Final rhoN}
\end{equation}

To investigate the behavior of the fitting formula ~(\ref{eqn:Final rhoN}) at large and small scattering angles, we insert the series  ~(\ref{eqn:Ei expansion origin}) and  ~(\ref{eqn:Ei expansion}) into the fitting formula and take the limit $t\to0$ , $t\to \infty$. 

For $\phi \ll \phi_0$, i.e. $t\to0$, we find
\begin{equation}
\begin{aligned}
 \lim_{t \to 0}\rho_N(t) = \frac{1}{2\pi \sigma^2}+\frac{1-\gamma-\ln \left ( \frac{\sigma^2}{2} \right )}{4\pi \sigma^4},
\end{aligned}
\end{equation}
For $\phi\gg\phi_0$, i.e. $t\to \infty$ we infer that
\begin{equation}
\begin{aligned}
 \lim_{t \to 0}\rho_N(t) &= -\frac{1}{4\pi \sigma^4}+\frac{1}{4\pi \sigma^4} \lim_{z \to \infty} \frac{z-1}{z} \left ( 1+\frac{1}{z}  + \frac{2}{z^2}+\frac{6}{z^3}+...\right ),\;\;z=\frac{t^2}{2\sigma^2}     \\
 &= \frac{1}{4\pi \sigma^4}\cdot \lim_{z \to \infty}  \frac{1}{z^2}
  =\frac{1}{4\pi \sigma^4}\cdot \lim_{t \to \infty} \frac{4\sigma^4}{t^4}\\
 &= \frac{1}{\pi t^4},
\end{aligned}
\end{equation}
and this result is generally consistent with that in K86.

In addition, the normalization of the fitting PDF given by Equation~(\ref{eqn:Final rhoN}) is required, and we prove that it has already been normalized to unity. This part is described in APPENDIX C.

We examine Equation~(\ref{eqn:Final rhoN}) by comparing its solution with numerical results of Equation~(\ref{eqn:neweq13}) and the simulations, as sketched in Fig.~\ref{fig:simulation_neweq}.
Solid lines are simulation results (left panel) and numerical results (right panel) of Equation~(\ref{eqn:neweq13}), while dashed lines are the solutions of Equation~(\ref{eqn:Final rhoN}) in both panels.
It can be concluded from the figure that the fitting PDF Equation~(\ref{eqn:Final rhoN}) is reliable, and its numerical solution is in good agreement with both the simulations and the numerical results of Equation~(\ref{eqn:neweq13}).

\begin{figure*}
	\centering                                                                   
	\includegraphics[width=\linewidth]{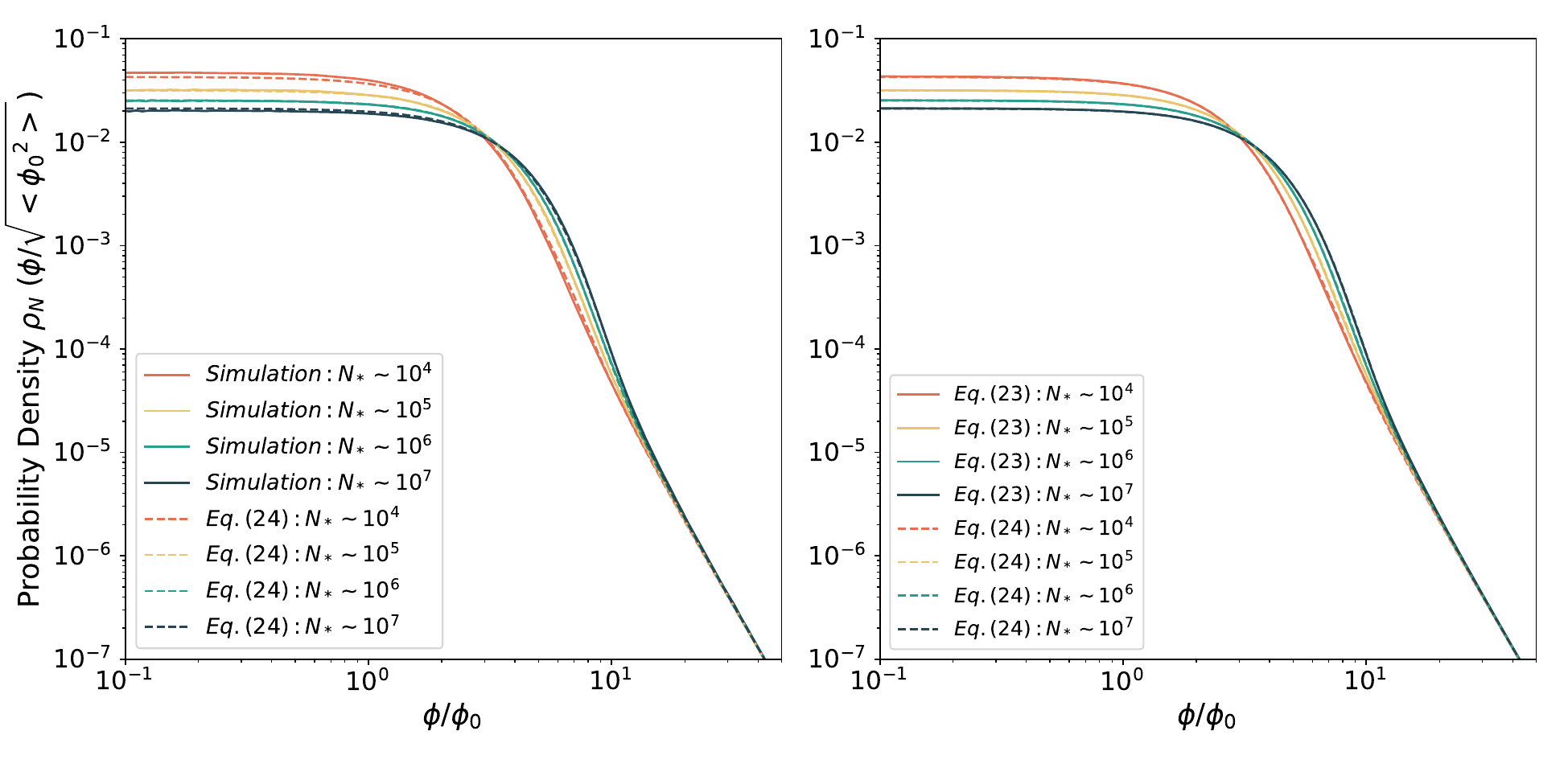}
    \caption{The probability density distribution of the deflection angle with a Salpeter mass distribution. Solid lines represent the distribution realized by our simulation (left) and the numerical result of Equation~(\ref{eqn:non_uni_rhoN}) (right), while dashed lines signify the numerical result of Equation~(\ref{eqn:fitting_non_uni_rhoN}). The star mass follows a Salpeter distribution, and the mass range is from $0.1M_\odot$ to $10M_\odot$, leading to $<M>=0.3M_\odot$. Different colors stand for different numbers of stars: $10^4$, $10^5$, $10^6$, $10^7$. 
    }
    \label{fig:Sal_simulation_neweq}
\end{figure*}

Furthermore, we expand our result to microlenses with a mass spectrum, along the path of \cite{1986ApJ...306....2K}. The PDF for a non-uniform mass distribution could be described as
\begin{equation}
\small
\begin{aligned}
    \rho_N(\phi) &= \frac{1}{2\pi} \int_{0}^{\infty}\mathrm{J_0}(k\phi)\tilde\rho_N(k)k\mathrm{d}k\\
    &= \frac{1}{2\pi<\phi_0^2>} \int_{0}^{\infty}\mathrm{J_0}\left ( \frac{x\phi}{\sqrt[]{<\phi_0^2>} } \right ) \tilde\rho_N(x)x\mathrm{d}x\\
    &= \frac{1}{2\pi<\phi_0^2>} \int_{0}^{\infty}\mathrm{J_0}
    \left ( \frac{x\phi}{\sqrt[]{<\phi_0^2>} } \right ) \times \exp\left \{ -\frac{x^2}{2} \ln\left [ \frac{1.454 f\sqrt{N} }{x}  \right ]  \right \}   x\mathrm{d}x,
\end{aligned} 
\label{eqn:non_uni_rhoN}
\end{equation}
where $f=\sqrt{<M^2>}\exp \left ( -\frac{<M^2\ln M>}{M^2} \right ) $.
The form of its fitting formula remains unchanged compared to Equation~(\ref{eqn:Final rhoN})
\begin{equation}
\begin{aligned}
    \rho_N(t)\simeq &\frac{e^{-z}}{2\pi \sigma^2}+\frac{1}{4\pi \sigma^4}\left [ 2e^{-z}-1 
     -(1-z)e^{-z}\left ( \mathrm{Ei}(z) -\ln z-\ln \frac{2}{\sigma^2} \right ) \right ],
     \;\;z=\frac{t^2}{2\sigma^2} \\
              =&   \frac{1}{2\pi \sigma^2}\exp{ \left(-\frac{t^2}{2\sigma^2}\right)} + \frac{1}{4\pi \sigma^4}\left \{ 2\exp{ \left(-\frac{t^2}{2\sigma^2}\right)} -1 \right. \\
              & -\left(1-\frac{t^2}{2\sigma^2}\right)\exp{ \left(-\frac{t^2}{2\sigma^2}\right)} \left [ \mathrm{Ei}\left ( \frac{t^2}{2\sigma ^2} \right )  \left. -\ln \left ( \frac{t^2}{2\sigma^2}\right ) -\ln \left ( \frac{2}{\sigma^2}\right ) \right ]  \right \},
\end{aligned}    
\label{eqn:fitting_non_uni_rhoN}
\end{equation}
except for $t=\frac{\phi}{\sqrt{<\phi_0^2>}}$, $\phi_0=\frac{M}{<M>}\sqrt{\kappa_*}$, and $\sigma^2=\ln(1.454 f\sqrt{N})$ .

According to the convention above, we also verify the validity of Equation~(\ref{eqn:non_uni_rhoN}) and Equation~(\ref{eqn:fitting_non_uni_rhoN}) similar to Fig.~\ref{fig:simulation_neweq},  as presented in Fig.~\ref{fig:Sal_simulation_neweq}. A Salpeter mass distribution with the mass range from $0.1M_\odot$ to $10M_\odot$ resulting in $<M>=0.3M_\odot$ is chosen in our realization. Notably, here in Fig.~\ref{fig:Sal_simulation_neweq} the deflection angle is scaled with $\sqrt{<\phi_0^2>}$. Good consistency can be drawn from both the left panel and the right panel indicating the reliability of Equation~(\ref{eqn:non_uni_rhoN}) and Equation~(\ref{eqn:fitting_non_uni_rhoN}).

\begin{figure*}
        \centering
	\includegraphics[width=12cm]{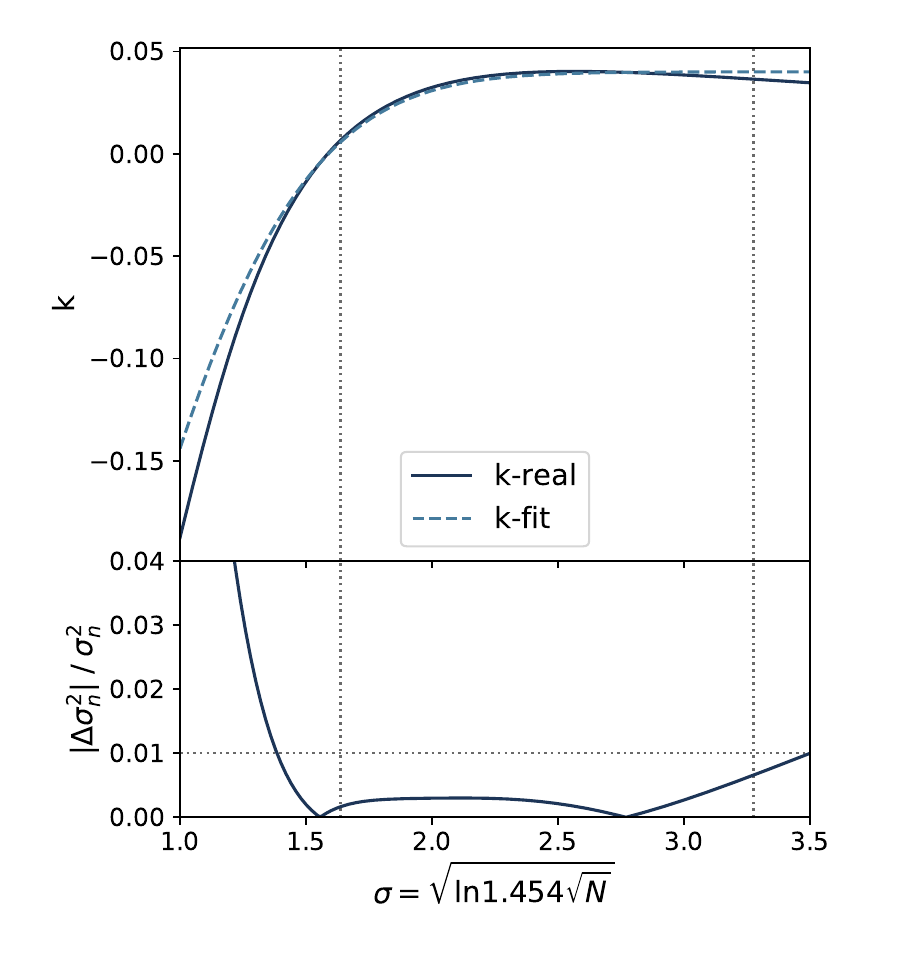}
    \caption{The comparison of the approximate $k$ and the real $k$ (top) and the relative error of $\sigma^2_n$ (bottom). Two vertical gray lines signify the value of $\sigma$ when $N=10^2$ and $N=10^9$, respectively. The horizontal gray line in the bottom panel marks the relative error of $\sigma^2_n$ equaling $1\%$.
    }
    \label{fig:Error_sigma_k_sigman}
\end{figure*}

\subsection{Gaussian approximation of the fitting formula}\label{sec:Gaussian approximation}
In the previous section, we obtained a fitting formula for the integral form of the probability distribution for microlensing induced deflection angle.
Nevertheless,  in order to simplify the following theoretical derivations, we aim to find a suitable Gaussian approximation, which is suggested by the central limit theorem applied to multiple deflection distributions that are dominated by small angles. 
Given the deviations between Equation~(\ref{eqn:Katz}) and our fitting formula and the simulations mentioned before, the Gaussian distribution obtained by K86 may not be the best approximation for our fitting formula.

Assuming that the standard deviation of the new Gaussian distribution is $\sigma_n$, we can use Chi-squared as a  measure to find a suitable $\sigma_n$ for the fitting formula ~(\ref{eqn:Final rhoN}).

We definr $\chi^2$ as
\begin{equation}
\chi ^2 = \int_{0}^{\infty } [\mathrm{G}(t,\sigma_n) - \rho_N(t)]^2 \cdot 2\pi t\mathrm{d}t,
\end{equation}
where $\mathrm{G}(t,\sigma_n) = \frac{1}{2\pi \sigma^2_n}e^{-\frac{t^2}{2\sigma^2_n}}$ stands for the new Gaussian distribution with $\sigma_n$ as an undetermined parameter. Then we take the partial derivative of $\sigma_n$ and make it equal to zero
\begin{equation}
\begin{aligned}
   &\frac{\partial \chi ^2}{\partial \sigma_n}  = 2\int_{0}^{\infty}  [\mathrm{G}(t,\sigma_n) - \rho_N(t)]\cdot \frac{\partial \mathrm{G}(t,\sigma_n)}{\partial \sigma_n }  \cdot 2\pi t\mathrm{d}t  =0 \\    
   &\Rightarrow \int_{0}^{\infty}  \mathrm{G}(t,\sigma_n) \cdot \frac{\partial \mathrm{G}(t,\sigma_n)}{\partial \sigma_n }  \cdot 2\pi t\mathrm{d}t  \\
   &=  \int_{0}^{\infty}  \rho_N(t)  \cdot \frac{\partial \mathrm{G}(t,\sigma_n)}{\partial \sigma_n }  \cdot  2\pi t\mathrm{d}t,
\end{aligned}   
\label{eqn:chi_square}
\end{equation}

From solving the integrals on both sides of the equation we acquire
\begin{equation}
\small
\begin{aligned}
    \frac{\sigma^6+3\sigma^4\sigma^2_n+(4\gamma-6-\sigma^2)\sigma^4_n-3\sigma^6_n-4\sigma^4_n \ln\left ( \frac{2}{\sigma^2+\sigma^2_n}  \right ) }{(\sigma^2+\sigma^2_n)^3} =0.
\end{aligned}    
\end{equation}

\begin{figure*}
        \centering
	\includegraphics[width=12cm]{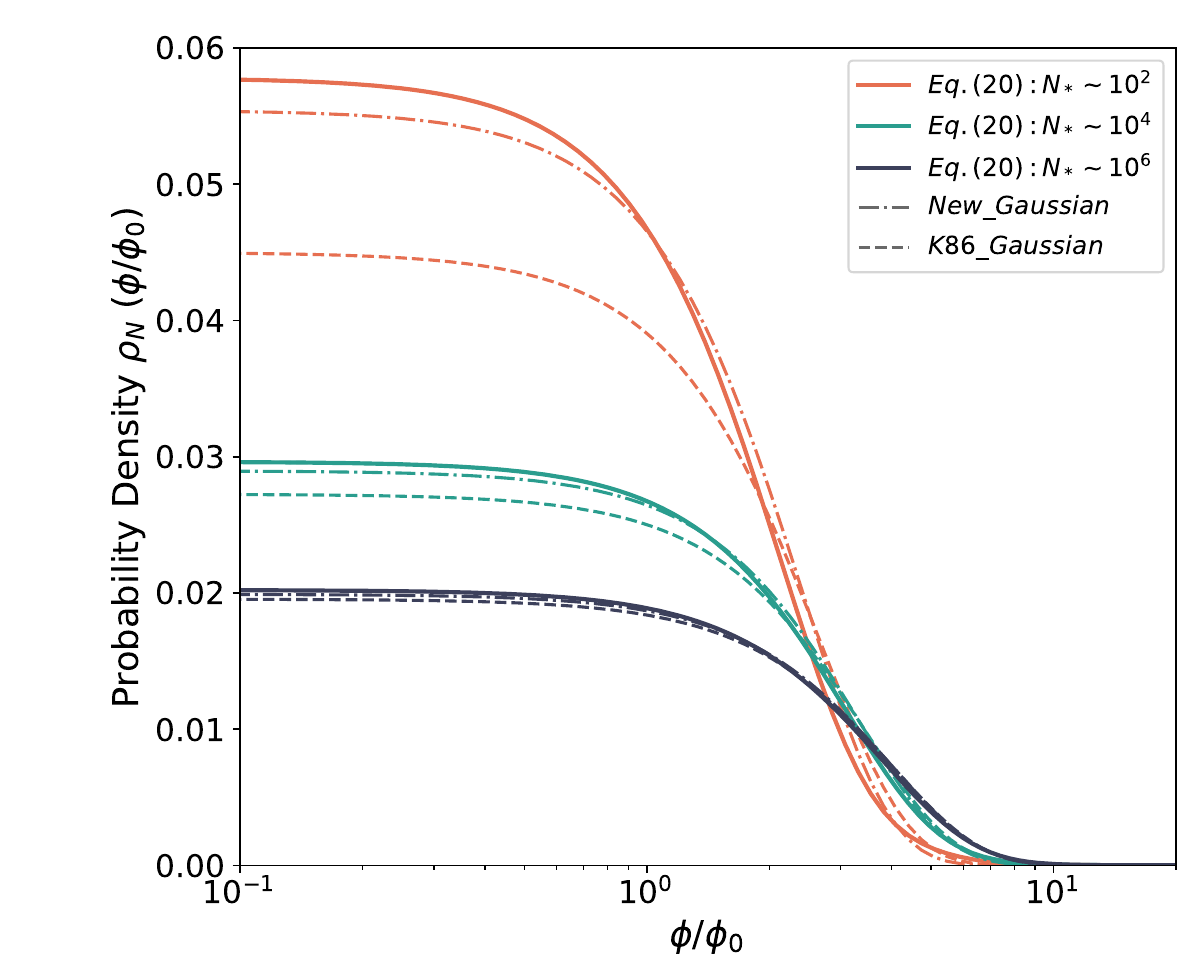}
    \caption{The comparison results of the fitting formula Equation~(\ref{eqn:Final rhoN}),  the new Gaussian distribution, and the Gaussian distribution from K86, the x-axis is logarithmic. The red, the green and the blue lines stand for the situation with $10^2$, $10^4$ and $10^6$ stars in the uniform mass. 
    }
    \label{fig:Gasussian_new}
\end{figure*}

Setting $\frac{\sigma^2_n-\sigma^2}{\sigma^2}=2k$, where $k\ll1$, then expanding $\ln(1+k)$ as $k-k^2/2$ and omitting higher order terms, we obtain a cubic equation in $k$
\begin{equation}
\begin{aligned}
    &2(\gamma+\ln\sigma^2)(1+2k)^2 + 4(1-3\sigma^3)k^3 -5(1+4\sigma^2)k^2 -2(5+4\sigma^2)k-3=0. 
\end{aligned}  
\end{equation}
Despite a complex solution for this cubic equation having been obtained, we decide to replace it with an approximate but more concise form 
\begin{equation}
    k\approx  0.04 - 0.5 e^{-\sigma^2},
\label{eqn:k_fit}
\end{equation}
then $\sigma^2_n$ can be written as $\sigma^2_n = (1.08-e^{-\sigma^2})\sigma^2$, whereby $\sigma^2=\ln(1.454\sqrt{N})$. It is expected that for a uniform mass distribution of stars, the $\sigma$ ranges from 1.63612 to 3.09591 when $N$ varies from $10^2$ to $10^8$.

How acceptable is it to take Equation~(\ref{eqn:k_fit}) as an  approximation for the real solution of $k$? In Fig.~\ref{fig:Error_sigma_k_sigman} we display the comparison of the approximate $k$ and the real $k$ in the top panel, and in the bottom one the relative error of $\sigma^2_n$ induced by the error of $k$ is presented. It can be seen from the figure that the relative error of $\sigma^2_n$ is within $1\%$ when $10^2<N<10^9$, noting that Equation~(\ref{eqn:k_fit}) is a reasonable approximation. 

Eventually, we obtain $\sigma_n$ as the standard deviation of the new Gaussian distribution. Fig.~\ref{fig:Gasussian_new},  presents the comparison result of the fitting formula ~(\ref{eqn:Final rhoN}),  the new Gaussian distribution and the Gaussian distribution with $\sigma^2=\ln(3.05\sqrt{N})$ from K86, showing that our new Gaussian distribution is in better agreement with the fitting formula ~(\ref{eqn:Final rhoN}).

\section{Summary}\label{sec:Summary}
The cosmological microlensing effect can be induced by the existence of compact objects in the lens surface density, leading to a variety of observable stochastic fluxes from the lensed sources. 
In order to improve our understanding of this effect and facilitate further analytical research, it is necessary to re-examine the microlensing deflection angle.
We calculate the PDF of the light deflection angle using a more appropriate assumption and obtain an integral expression that has better agreement with the simulation results. 
Moreover, through mathematical deduction, we obtain a fitting formula for this integral expression, which allows for the direct and almost accurate calculation of the deflection angle probability density distribution.
In addition, we find a more suitable Gaussian approximation for this fitting formula, which can simplify subsequent theoretical work.

\section*{Acknowledgements}

This work is supported by the National Natural Science Foundation of China (NSFC, Grant Nos. U1931210, 11673065, 11273061).
We acknowledge the cosmology simulation database (CSD) in the National Basic Science Data Centre (NBSDC) and its funds the NBSDC-DB-10 (No.2020000088)
We acknowledge the science research grants from the China Manned
Space Project with No.CMS-CSST-2021-A12.

\section*{Data Availability}

This theoretical study does not generate any new data.


\appendix
\section{The derivation of Equation (8)}

For Equation~(\ref{eqn:tilde_rho1_k}) at $a\ll$1, we make some deductions by integrating it by parts and using some basic properties of a Bessel function: 
\begin{equation}
\begin{aligned}
    \tilde\rho_1(k)=\tilde\rho_1(a)= &\int_{0}^{\infty} \mathrm{J}_0(ax)\left ( \frac{x^2+2}{\sqrt{x^2+4}}-x\right)\mathrm{d}x \\
                                   = &\left. \mathrm{J}_0(ax)\left(-\frac{x^2}{2}+\frac{x}{2}\sqrt{x^2+4}\right) \right| ^{\infty }_{0} -  a\int_{0}^{\infty} \mathrm{J}_1(ax)\left (\frac{x^2}{2}-\frac{x}{2}\sqrt{x^2+4}\right)\mathrm{d}x \\
                                   = &-a\int_{0}^{\infty} \mathrm{J}_1(ax)\left (\frac{x^2}{2}-\frac{x}{2}\sqrt{x^2+4}\right)\mathrm{d}x\\    
                                   = &-  a\int_{0}^{\infty}\mathrm{J}_1(ax)+ \mathrm{J}_1(ax)\left (\frac{x^2}{2}-\frac{x}{2}\sqrt{x^2+4}\right)\mathrm{d}x +a\int_{0}^{\infty}\mathrm{J}_1(ax)\mathrm{d}x \\  
                                   = &1-a\int_{0}^{\infty}\mathrm{J}_1(ax)\left ( 1+\frac{x^2}{2}-\frac{x}{2}\sqrt{x^2+4}\right )\mathrm{d}x,
\label{eqn:tilrho1}
\end{aligned}
\end{equation}
noting that F(0)=1. 

For small $a$, the Taylor series of the $\mathrm{J}_1$ state is
\begin{equation}
    \mathrm{J}_1(ax)=\frac{ax}{2}+O(a^3).
\label{eqn:expansion}
\end{equation}
Since expanding Equation~(\ref{eqn:expansion}) should be valid for some finite area, i.e. $0< ax< s$, we substitute the upper limit of integral Equation~(\ref{eqn:tilrho1}) by an undetermined parameter $\alpha=s/a$, which is expected to be much larger than unity for $a\ll1$. 
Thus we may expand Equation~(\ref{eqn:tilrho1}) on $a$
\begin{equation}
\begin{aligned}
    \tilde\rho_1(k)= &1-\frac{a^2}{2} \int_{0}^{\alpha}x\left ( 1+\frac{x^2}{2}-\frac{x}{2}\sqrt{x^2+4}\right )\mathrm{d}x+o(a^4)\\
                   = &1-\frac{a^2}{2}\left ( \frac{\alpha^2}{2} +\frac{\alpha^4}{8}-\frac{\alpha}{8}\left(\alpha^2+2\right)\sqrt{\alpha^2+4} + \mathrm{arcsinh}\left(\frac{\alpha}{2}\right)\right )+O(a^4).
\end{aligned}
\end{equation}
Then we expand the function of $\alpha$ near infinity
\begin{equation}
\begin{aligned}
&\lim_{\alpha \to \infty }\left(\frac{\alpha^2}{2}+\frac{\alpha ^4}{8} -\frac{\alpha}{8}\left(\alpha^2+2\right)\sqrt{\alpha^2+4}\right) = -\frac{1}{4},\\
&\mathrm{arcsinh}\left(\frac{\alpha}{2}\right)\sim \ln\alpha + O(\alpha^{-2}),
\end{aligned}    
\end{equation}


thus
\begin{equation}
    \tilde\rho_1(k)=1-\frac{a^2}{2} \left ( \ln\frac{s}{a}-\frac{1}{4}\right ) + O(a^4), \;\; a\ll 1.
\end{equation}

Theoretically, we expect the value of $s$ should be very close to 1.8412, which is the first root of transcendental equation $\mathrm{J}_0(x)=\mathrm{J}_2(x)$. Further numerical test suggests that $s\simeq1.866$ is a better choice, which gives
\begin{equation}
\begin{aligned}
    \tilde\rho_1(k) &\simeq 1-\frac{a^2}{2} \ln\left(\frac{1.866}{a\sqrt[4]{e}} \right)
    \simeq  1-\frac{a^2}{2} \ln\left(\frac{1.454}{a}\right)\\
    &\simeq  \exp\left [ -\frac{a^2}{2}\ln\left({\frac{1.454}{a}}\right)\right].
\end{aligned}
\end{equation}

\section{The derivation of Equation (15)}
The first integral formula of Equation~(\ref{eqn:rhont 2 integral}) is a special case $(l=1)$ of the following result
\begin{equation}
\begin{aligned}
     &\int_{0}^{\infty}x^l\mathrm{J}_0(xt)\exp\left ( - \frac{\sigma^2x^2}{2} \right)\mathrm{d}x\\
     =&\frac{1}{\sigma^2}\left ( \frac{2}{\sigma^2}\right )^{(l-1)/2}\sum_{n=0}^{\infty}\frac{\Gamma (n+\frac{l+1}{2} )}{(n!)^2}\left ( -\frac{t^2}{2\sigma^2}\right )^n,   
\label{eqn:integral_l}
\end{aligned}
\end{equation}
Equation~(\ref{eqn:integral_l}) can be derived by inserting the Taylor series of $\mathrm{J}_0$ into the left integral, and exchanging the integration and summation. 

For $l=1$, we have
\begin{equation}
\begin{aligned}
     \int_{0}^{\infty}x \mathrm{J}_0(xt)\exp\left ( - \frac{\sigma^2x^2}{2} \right)\mathrm{d}x&=\frac{1}{\sigma^2} \sum_{n=0}^{\infty}\frac{\Gamma (n+1)}{(n!)^2}\left ( -\frac{t^2}{2\sigma^2}\right )^n\\
     &=\frac{1}{\sigma^2} \exp{ \left(-\frac{t^2}{2\sigma^2}\right)}.
\label{eqn:integral1}
\end{aligned}
\end{equation}

To deal with the second term in Equation~(\ref{eqn:rhont 2 integral}), we take the derivative of $l$ on both sides of Equation~(\ref{eqn:integral_l}) and let $l=3$,

\begin{equation}
\begin{aligned}
&\int_{0}^{\infty}x^3\ln x\mathrm{J}_0(xt)\exp\left ( - \frac{\sigma^2x^2}{2} \right)\mathrm{d}x\\
   =& \frac{1}{\sigma^2}  \left[ \left ( \frac{2}{\sigma^2}  \right )^{(l-1)/2}  \ln\left ( \frac{2}{\sigma^2}\right )\cdot \frac{1}{2}\sum_{n=0}^{\infty}\frac{\Gamma(n+\frac{l+1}{2})}{(n!)^2} \left ( \frac{-t^2}{2\sigma^2} \right )^n + \left. \left ( \frac{2}{\sigma^2}  \right )^{(l-1)/2}\cdot \frac{\partial \left ( \sum_{n=0}^{\infty}\frac{\Gamma(n+\frac{l+1}{2})}{(n!)^2} \left ( \frac{-t^2}{2\sigma^2} \right )^n  \right ) }{\partial l} \right] \right| _{l=3} \\  
   =&  \frac{1}{\sigma^2} \cdot \left ( \frac{2}{\sigma^2}  \right )^{(l-1)/2} \left[   \ln\left ( \frac{2}{\sigma^2}\right )\cdot \frac{1}{2}\sum_{n=0}^{\infty}\frac{\Gamma(n+\frac{l+1}{2})}{(n!)^2} \left ( \frac{-t^2}{2\sigma^2} \right )^n +  \left. \frac{1}{2} \cdot \sum_{n=0}^{\infty } \frac{(H_{n+\frac{l+1}{2}-1}-\gamma)\Gamma(n+\frac{l+1}{2})}{(n!)^2}  \left(\frac{-t^2}{2\sigma^2} \right)^n    \right]  \right| _{l=3}    \\
   =&  \frac{1}{\sigma^4} \left[   \ln\left ( \frac{2}{\sigma^2}\right )\cdot \sum_{n=0}^{\infty}\frac{\Gamma(n+2)}{(n!)^2} \left ( \frac{-t^2}{2\sigma^2} \right )^n  +  \sum_{n=0}^{\infty } \frac{(H_{n+1}-\gamma)\Gamma(n+2)}{(n!)^2}  \left(\frac{-t^2}{2\sigma^2} \right)^n    \right]   \\
   =&\frac{1}{\sigma^4}\left[  \left ( \ln\frac{2}{\sigma^2}-\gamma  \right )\left ( 1-\frac{t^2}{2\sigma^2}\right )\exp \left( -\frac{t^2}{2\sigma^2}\right) \right.\left.+\sum_{n=0}^{\infty}\frac{(n+1 )}{n!}H_{n+1}\left ( -\frac{t^2}{2\sigma^2}\right )^n \right].
\end{aligned}
\end{equation}

\section{normalization of the fitting probability density function}
We will prove that the fitting PDF given by Equation~(\ref{eqn:Final rhoN}) has already been normalized to unity.
Here, we list two basic integral formulas,
\begin{equation}
\begin{aligned}
    &\int_{0}^{\infty }x \exp \left ( -\frac{x^2}{2\sigma^2}  \right )\mathrm{d}x = \sigma^2,\\  
    &\int_{0}^{\infty }x \left ( 1-\frac{x^2}{2\sigma^2} \right )  \exp \left ( -\frac{x^2}{2\sigma^2} \right)\mathrm{d}x = 0,
\end{aligned}    
\end{equation}

Combining them with the series ~(\ref{eqn:Ei expansion origin}), by straightforward practice we have
\begin{equation}
\small
\begin{aligned}
&\int_{0}^{\infty} 2\pi t\cdot \rho_N(t)\mathrm{d}t\\
=& \frac{1}{\sigma^2} \int_{0}^{\infty } t \exp\left (\frac{-t^2}{2\sigma^2} \right )  \mathrm{d}t + \frac{1}{\sigma^4} \int_{0}^{\infty } t \exp\left (\frac{-t^2}{2\sigma^2} \right )  \mathrm{d}t  \\
&- \frac{1}{2\sigma^4} \int_{0}^{\infty } \left \{  t+t \left(1-\frac{t^2}{2\sigma^2}\right)\exp\left (\frac{-t^2}{2\sigma^2} \right ) \left [\gamma +\sum_{n=1}^{\infty }\frac{\left ( \frac{t^2}{2\sigma^2} \right )  ^n}{n\cdot n!} -\ln \left ( \frac{2}{\sigma^2}\right ) \right ]  \right \} \mathrm{d}t   \\
=& 1+\frac{1}{\sigma^2}  - \frac{1}{4\sigma^4} \int_{0}^{\infty } \left [1+ \left(1-\frac{t^2}{2\sigma^2}\right)\exp\left (\frac{-t^2}{2\sigma^2} \right )  \sum_{n=1}^{\infty }\frac{\left ( \frac{t^2}{2\sigma^2} \right )  ^n}{n\cdot n!} \right ]   \mathrm{d}t^2, \\
 &\;\;let \; k=t^2, \;1= \lim_{p \to 0} e^{-pk}   \\
=&1+\frac{1}{\sigma^2}-\frac{1}{4\sigma^4}  \int_{0}^{\infty}\left[ e^{-pk}  + \left. \left ( 1- \frac{k}{2\sigma^2}\right)\exp\left (\frac{-t^2}{2\sigma^2}-pk \right ) \sum_{n=1}^{\infty}\frac{1}{n\cdot n!}\left ( \frac{k}{2\sigma^2}\right )^n \right] \mathrm{d}k  \right| _{p\to 0}  \\
=&1+\frac{1}{\sigma^2}-\frac{1}{4\sigma^4}  \left[ \frac{1}{p} + \left.   \sum_{n=1}^{\infty}\frac{1}{n\cdot n!\left(2\sigma^2\right)^n}\int_{0}^{\infty} k^n\left ( 1- \frac{k}{2\sigma^2}\right) \exp\left (\frac{-t^2}{2\sigma^2}-pk \right ) \mathrm{d}k\right]   \right| _{p\to 0}  \\
=&1+\frac{1}{\sigma^2}-\frac{1}{4\sigma^4} \left.  \left [\frac{1}{p} - \sum_{n=1}^{\infty}\frac{2\sigma^2(n-2\sigma^2p)}{n\left(1+2\sigma^2p\right)^{n+2} } \right ]   \right| _{p\to 0}\\
=&1+\frac{1}{\sigma^2}-\frac{1}{4\sigma^4} \left.  \left [\frac{1}{p} - \frac{1+4\sigma^4p^2\ln \left ( \frac{2\sigma^2p}{1+2\sigma^2p}  \right ) }{p\left(1+2\sigma^2p\right)^2 } \right ]   \right| _{p\to 0}\\
=&1+\frac{1}{\sigma^2}-\frac{1}{4\sigma^4} \cdot 4\sigma^2=1.
\end{aligned}
\end{equation}

\bibliographystyle{raa}
\bibliography{references}  

\begin{thebibliography}{29}
\providecommand\natexlab[1]{#1}
\providecommand\JournalTitle[1]{#1}

\bibitem[{Chang} \& {Refsdal}(1979)]{1979Natur.282..561C}
{Chang}, K., \& {Refsdal}, S. 1979, \nat, 282, 561

\bibitem[{Chen} {et~al.}(2019)]{2019ApJ...881....8C}
{Chen}, W., {Kelly}, P.~L., {Diego}, J.~M., {et~al.} 2019, \apj, 881, 8

\bibitem[{Dai} \& {Pascale}(2021)]{2021arXiv210412009D}
{Dai}, L., \& {Pascale}, M. 2021, arXiv e-prints, arXiv:2104.12009

\bibitem[{Diego}(2022)]{2022A&A...665A.127D}
{Diego}, J.~M. 2022, \aap, 665, A127

\bibitem[{Diego} {et~al.}(2022)]{2022A&A...665A.134D}
{Diego}, J.~M., {Pascale}, M., {Kavanagh}, B.~J., {et~al.} 2022, \aap, 665,
  A134

\bibitem[{Irwin} {et~al.}(1989)]{1989AJ.....98.1989I}
{Irwin}, M.~J., {Webster}, R.~L., {Hewett}, P.~C., {Corrigan}, R.~T., \&
  {Jedrzejewski}, R.~I. 1989, \aj, 98, 1989

\bibitem[{Jim{\'e}nez-Vicente} \& {Mediavilla}(2019)]{2019ApJ...885...75J}
{Jim{\'e}nez-Vicente}, J., \& {Mediavilla}, E. 2019, \apj, 885, 75

\bibitem[{Katz} {et~al.}(1986)]{1986ApJ...306....2K}
{Katz}, N., {Balbus}, S., \& {Paczynski}, B. 1986, \apj, 306, 2

\bibitem[{Kaurov} {et~al.}(2019)]{2019ApJ...880...58K}
{Kaurov}, A.~A., {Dai}, L., {Venumadhav}, T., {Miralda-Escud{\'e}}, J., \&
  {Frye}, B. 2019, \apj, 880, 58

\bibitem[{Kayser} {et~al.}(1986)]{1986A&A...166...36K}
{Kayser}, R., {Refsdal}, S., \& {Stabell}, R. 1986, \aap, 166, 36

\bibitem[{Kelly} {et~al.}(2018)]{2018NatAs...2..334K}
{Kelly}, P.~L., {Diego}, J.~M., {Rodney}, S., {et~al.} 2018, Nature Astronomy,
  2, 334

\bibitem[{Kochanek}(2004)]{2004ApJ...605...58K}
{Kochanek}, C.~S. 2004, \apj, 605, 58

\bibitem[{Lewis} {et~al.}(1993)]{1993MNRAS.261..647L}
{Lewis}, G.~F., {Miralda-Escude}, J., {Richardson}, D.~C., \& {Wambsganss}, J.
  1993, \mnras, 261, 647

\bibitem[{Mediavilla} {et~al.}(2006)]{2006ApJ...653..942M}
{Mediavilla}, E., {Mu{\~n}oz}, J.~A., {Lopez}, P., {et~al.} 2006, \apj, 653,
  942

\bibitem[{Mediavilla} {et~al.}(2009)]{2009ApJ...706.1451M}
{Mediavilla}, E., {Mu{\~n}oz}, J.~A., {Falco}, E., {et~al.} 2009, \apj, 706,
  1451

\bibitem[{Meena} {et~al.}(2022)]{2022MNRAS.514.2545M}
{Meena}, A.~K., {Arad}, O., \& {Zitrin}, A. 2022, \mnras, 514, 2545

\bibitem[{Mortonson} {et~al.}(2005)]{2005ApJ...628..594M}
{Mortonson}, M.~J., {Schechter}, P.~L., \& {Wambsganss}, J. 2005, \apj, 628,
  594

\bibitem[{Rodney} {et~al.}(2018)]{2018NatAs...2..324R}
{Rodney}, S.~A., {Balestra}, I., {Bradac}, M., {et~al.} 2018, Nature Astronomy,
  2, 324

\bibitem[{Schechter} {et~al.}(2014)]{2014ApJ...793...96S}
{Schechter}, P.~L., {Pooley}, D., {Blackburne}, J.~A., \& {Wambsganss}, J.
  2014, \apj, 793, 96

\bibitem[{Schneider} \& {Wei{\ss}}(1986)]{1986MPARp.234.....S}
{Schneider}, P., \& {Wei{\ss}}, A. 1986, Max Planck Institut fur Astrophysik
  Report, 234

\bibitem[{Shalyapin} {et~al.}(2021)]{2021A&A...653A.121S}
{Shalyapin}, V.~N., {Gil-Merino}, R., \& {Goicoechea}, L.~J. 2021, \aap, 653,
  A121

\bibitem[{Sluse} {et~al.}(2012)]{2012A&A...544A..62S}
{Sluse}, D., {Hutsem{\'e}kers}, D., {Courbin}, F., {Meylan}, G., \&
  {Wambsganss}, J. 2012, \aap, 544, A62

\bibitem[{Thompson} {et~al.}(2010)]{2010NewA...15...16T}
{Thompson}, A.~C., {Fluke}, C.~J., {Barnes}, D.~G., \& {Barsdell}, B.~R. 2010,
  \na, 15, 16

\bibitem[{Venumadhav} {et~al.}(2017)]{2017ApJ...850...49V}
{Venumadhav}, T., {Dai}, L., \& {Miralda-Escud{\'e}}, J. 2017, \apj, 850, 49

\bibitem[{Wambsganss}(1999)]{1999JCoAM.109..353W}
{Wambsganss}, J. 1999, Journal of Computational and Applied Mathematics, 109,
  353

\bibitem[{Welch} {et~al.}(2022)]{2022Natur.603..815W}
{Welch}, B., {Coe}, D., {Diego}, J.~M., {et~al.} 2022, \nat, 603, 815

\bibitem[{Witt}(1993)]{1993ApJ...403..530W}
{Witt}, H.~J. 1993, \apj, 403, 530

\bibitem[{Wyithe} \& {Turner}(2001)]{2001MNRAS.320...21W}
{Wyithe}, J.~S.~B., \& {Turner}, E.~L. 2001, \mnras, 320, 21

\bibitem[{Zheng} {et~al.}(2022)]{2022ApJ...931..114Z}
{Zheng}, W., {Chen}, X., {Li}, G., \& {Chen}, H.-Z. 2022, \apj, 931, 114

\end{thebibliography}

\end{document}